# CEP-stable high-energy ytterbium doped fiber amplifier

M. NATILE[1,2]*, L. LAVENU[1,3], F. GUICHARD[1], M. HANNA[3], Y. ZAOUTER[1], R. CHICHE[4], X. CHEN[1], J.F. HERGOTT[2], W. BOUTU[2], H. MERDJI[2] AND P. GEORGES[3]

1. *Amplitude Laser Group, 2 – 4 rue du Bois Chaland CE 2926, 91029 Evry, France*
2. *LIDYL, CEA, CNRS, Université Paris-Saclay, UMR 9222 CEA-SACLAY, 91191 Gif-sur-Yvette, France*
3. *Laboratoire Charles Fabry, Institut d'Optique Graduate School, CNRS, Université Paris-Saclay, 91127 Palaiseau Cedex, France*
4. *Laboratoire de l'Accélérateur Linéaire, IN2P3, CNRS, Université Paris-Saclay, 91898 Orsay Cedex, France*

*Corresponding author: michele.natile@amplitude-laser.com*

*We report on the CEP stabilization of an Yb-doped fiber amplifier system delivering 30 µJ, 100 fs pulses at 100 kHz repetition rate. A single shot, every shot, measurement of the CEP stability based on a simple f-2f interferometer is performed, yielding a CEP standard deviation of 320 mrad rms over 1 s. Long-term stability is also assessed, with 380 mrad measured over one hour. This level of performance is allowed by a hybrid architecture including a passively CEP-stabilized front-end based on difference frequency generation, and an active CEP stabilization loop for the fiber amplifier system, acting on a telecom-grade integrated LiNbO3 phase modulator. These results demonstrate the full compatibility of Yb-doped high repetition rate laser for attoscience.*

Carrier-envelope phase (CEP) stabilized few-cycle amplified laser systems are one of the keys enabling technique for attoscience [1, 2]. Since the first demonstration of isolated attosecond pulse generation [3], applications requiring a precise control of the electric field have considerably expanded. To date, titanium-doped sapphire (Ti:Sa) amplifier systems have mostly provided the required short and energetic pulses, at repetition rates ranging from 50 Hz to 10 kHz [4, 5].

A large number of applications would benefit from further repetition rate scaling to ≥ 100 kHz reducing acquisition times and improving signal-to-noise ratio. This is particularly true for low yield experiments involving coincidence detection [6, 7]. At repetition rates above 10 kHz, high power Ti:Sa amplifiers become complicated, costly and do not provide the long term stability and robustness required by today's science. A proposed and demonstrated alternative is using CEP-stable Ti:Sapphire oscillator to seed high-repetition rate optical parametric chirped pulse amplifiers (OPCPA) [8, 9]. Of particular interest, OPCPA can be configured to work at different wavelengths opening new opportunities. Unfortunately, despite remarkable demonstrations, power scaling has been shown hiding unexpected difficulties. OPCPA is, furthermore, fundamentally limited by the low conversion efficiency from the pump to the useable beam (~10%) before spatial and temporal distortions occur. Thus, very powerful, hence costly, pump sources have to be employed to generate only a few tens of Watts [10]. In the past years, laser sources based on Ytterbium gain material have shown stunning performances with >kW average power and >mJ energy per pulses [11, 12]. Industrial Ytterbium solid-state and/or fiber amplifier products are rapidly penetrating medical and industrial markets owing to their inherent performances, superior long-term robustness and reduced cost of ownership. Furthermore, efficient non-linear compression of these laser sources down to the few cycle regime has been demonstrated [13, 14] and used to drive high-photon flux XUV sources [15, 16] through high harmonic generation (HHG). However, robust and performant CEP stabilization is still missing. This is due to a number of reasons such as noisy fiber oscillator characteristics [17], large stretching and compression ratio and enhanced intensity to CEP noise transfer in different nonlinear stages [18]. To date, CEP-stabilization of Ytterbium-based amplified systems have been realized in linear amplification regime at µJ level with a regenerative Yb:KGW system at 1 MHz seeded by a solid state Yb:KGW oscillator [19] and with a fiber amplifier at 80 MHz seeded by a CEP-stable Ti:Sa oscillator [20]. Both systems revealed CEP fluctuations comparable to Ti:Sa and OPCPA sources, but are not directly scalable in average power for the former one [19] and in pulse energy for the latter one [20]. A CEP-stable Tm-doped linear fiber CPA architecture at 2 µm has also been demonstrated recently [21].

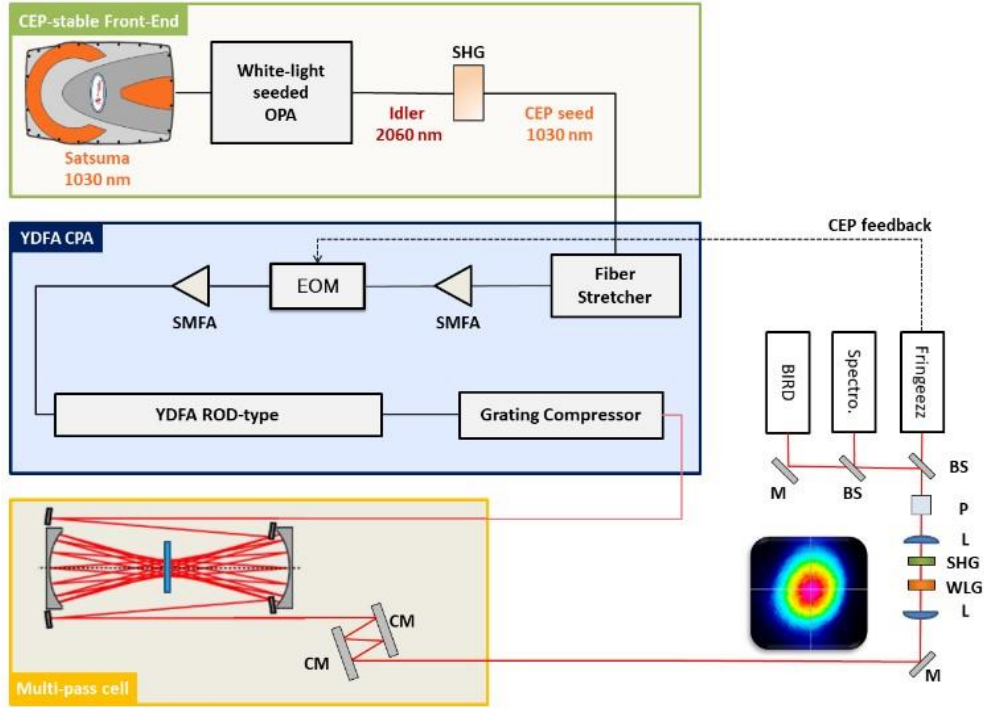

*Fig. 1. Experimental setup. SHG: Second Harmonic Generation, WLG: White Light Generation, OPA: Optical Parametric Amplification, SMFA: Single-mode fiber amplifier, EOM: Electro-optical modulator, CM: Chirped Mirrors, M: Mirror, L: lens, P: Polarizer, BS: Beam splitter.*

In this letter, we report on the first high-energy CEP-stabilized Yb-doped fiber chirped-pulse amplifier (FCPA) system, including a multi-pass cell (MPC) nonlinear compression stage [22]. It delivers 30 µJ 96 fs pulses at 100 kHz, and relies on several key elements:
- A passively CEP stable front-end at a central wavelength of 1030 nm.
- A FCPA system including preamplifiers, rodtype fiber power amplifier, and a large stretching/compression ratio.
- An active CEP feedback loop including an in-line in-focus f-to-2f interferometer and an integrated electro-optic phase modulator as an actuator.

The CEP stability is characterized in detail both at the full 100 kHz bandwidth over 1 s and at 10 kHz over 1 h, revealing < 400 mrad CEP stability on a shot–to shot basis. This is, to our knowledge, the first spectral f-2f CEP measurement ever reported at a repetition rate > 10 kHz. Overall, this source demonstrates that Yb-based ultrafast laser technology is fully compatible with CEP stabilization, paving the way for a future generation of compact and efficient laser drivers for attoscience applications.

The experiment is depicted in Figure 1. It consists of a passive CEP-stable front-end, seeding a large stretching/compression ratio high energy FCPA, followed by a MPC temporal compression stage.

The frontend uses a commercially available industrial FCPA (Satsuma, Amplitude Laser) to pump an optical parametric amplifier (OPA). Both white-light generated signal and OPA pump originates from the same laser pulse train. In this configuration, the idler radiation is inherently CEP-stabilized [23]. Here, the idler wavelength is tuned at 2060 nm and further converted to 1030 nm through second harmonic generation. Up to 400 nJ of energy per pulse is produced and seeded in the FCPA with a full width at half maximum (FWHM) pulse duration of 185 fs. The repetition rate of the CEP-stable seeder is furthermore tunable from single shot to 500 kHz through the control of the OPA pump laser repetition rate.

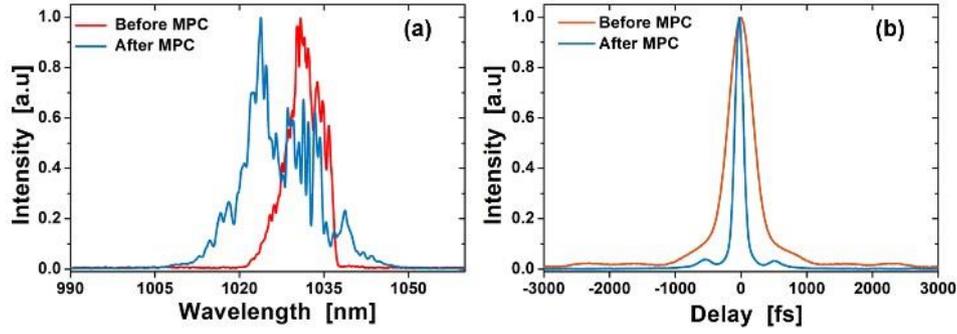

*Fig. 2. Spectra (a) and autocorrelation traces (b) at the output of the CPA (red) and MPC-based temporal compression (blue).*

The high energy FCPA is composed of a large stretching ratio fiber stretcher (pulse duration ~500 ps), followed by two single mode low power fiber pre-amplifiers from either side of an integrated LiNbO$_3$ phase modulator (PM) used as CEP control device. The possibility to stabilize the CEP of a Ti:Sapphire amplified laser using a bulk electro-optic modulator was demonstrated in [23]. The voltage required to shift the CEP phase by $\pi$ is related to the usual V$_\pi$ by

$$V_{\pi,CEP} = V_\pi \frac{n_e r_{33}}{\lambda\left(3r_{33}\frac{\partial n_e}{\partial \lambda}+n_e\frac{\partial r_{33}}{\partial \lambda}\right)}, \quad (1)$$

where $\lambda$ is the wavelength, $n_e$ is the extraordinary refractive index of LiNbO$_3$ and $r_{33}$ is one component of its electro-optic tensor. We estimate the value of the right-hand side fraction in Eq. 1 to be 7, which is coherent with the experimentally measured value of $V_{\pi,CEP}$ of 8 V, and the specified V$_\pi$ <2V. The power amplifier is composed of a 1-m long rod-type fiber with a mode field diameter of 65 µm pumped by a high power 976 nm diode laser. Finally, a highly dispersive 1750 l/mm grating compressor is setup to compress the pulse down to 340 fs.

Additional temporal compression is performed by means of a bulk MPC. The MPC is made of two spherical mirrors with a radius of curvature of 200 mm, separated by 240 mm. Pulses recirculate for 15 roundtrips in the MPC in which a 2.3 mm-thick antireflection coated silica plate is inserted. Here the role of the MPC is twofold. First, the shorter pulse width increases the coherence of the f-2f white light generation process for CEP measurement. Second, it is included as a test bench to study CEP stability in the presence of a highly nonlinear compression stage that must be included in a future high power source for attosecond physics. Figure 2 shows the spectral and temporal characteristics of the pulses at the output of the FCPA and after the MPC. At the maximum applicable input energy to the MPC i.e. 30 µJ, the spectrum is broadened from 16 nm to 45 nm leading to a pulse duration of 96 fs out of the MPC (assuming a Gaussian shape for the deconvolution), after chirp removal using 4800 fs² of negative dispersion. Finally, an in-line, in-focus f-to-2f interferometer setup is used for CEP characterization [25]. A pulse energy of ~1.5 µJ is sampled out of the MPC beam and focused into a 3 mm-thick

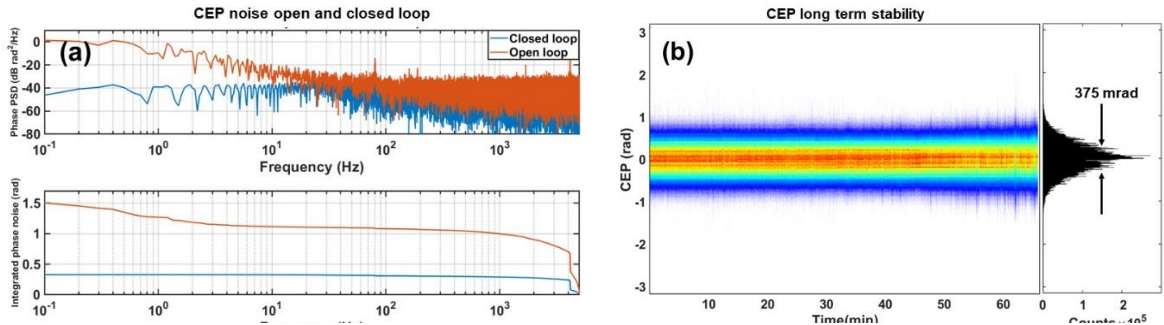

*Fig. 3. In loop single-shot CEP characterization at 10 kHz sampling rate (a) CEP PSD and IPN for open and closed loop operations. (b) Left: CEP density plot in closed loop over 60 minutes. Right: Corresponding CEP histogram.*

YAG crystal for octave-spanning spectral broadening. In the same focus, a 2 mm-thick BBO crystal is used to frequency double the spectral content in the 1100 – 1150 nm range, generating the spectral interference pattern in the 550 – 575 nm range after a polarizing cube.

We now focus on the in-depth characterization of the CEP stability. A commercial fast CEP measurement device (Fringeezz, *Fastlite*) is used at the output of the f-to-2f interferometer to acquire single shot CEP values at 10 kHz sampling rate. This signal is fed to a digital PID servo-controller that drives the phase modulator. Figure 3(a) shows the in-loop CEP power spectral density (PSD) and RMS integrated phase noise (IPN) in open and closed loop operation, acquired over 10 s. In the closed loop case, an overall IPN of 320 mrad is obtained, owing to the strong reduction of low frequency noise allowed by the feedback loop. A long-term CEP density plot (one histogram per second) over 1 h is also shown in Fig. 3(b) and yields an overall IPN of 375 mrad, demonstrating the robustness of the system.

Out-of-loop characterization is performed by measuring the spectral fringes using a fast spectrometer (AvaSpecs-ULS3648, *Avantes*). The minimum integration time of 30 µs allows to measure CEP fluctuations integrating only over 3 pulses, at a maximum sampling rate of 330 Hz.

Figure 4(a) shows the fringe spectrum acquired on the first acquisition (used to measure the fringe visibility), Figure 4(b) shows the fringe pattern as a function of time, Figure 4(c) shows the corresponding CEP variation as a function of time [26], and Figure 4(d) shows the overall CEP histogram over 3 s for three different integration times of 30 µs, 100 µs and 1 ms, corresponding respectively to 3, 10, and 100 shots averaging. The IPN increases from 56 mrad to 115 mrad and 214 mrad for decreasing integration times, with corresponding noise bandwidths of [0.3 Hz – 1 kHz], [0.3 Hz - 10kHz], and [0.3 Hz – 33 kHz]. An interesting point is to compare the measured fringe contrast, which is sometimes used as a way to characterize the CEP stability [27]. In our case, it remains essentially constant, with a visibility increasing from 68.3% to 68.9% from the less to the most integrated case. This visibility is therefore not a meaningful indication of CEP stability in these conditions.

Another out-of-loop CEP measurement is also carried out, with a focus on measuring the complete PSD function with no downsampling, at the full repetition rate, using the Beat Interferometer for Rapid Detection (BIRD, Amplitude Laser) technique [28]. This measurement is completely analog and based on the use of the difference signal from two photomultipliers detecting the signals of a spatially-dispersed spectral fringe hitting the apex of a prism. Figure 5(a) shows the closed loop PSD and corresponding IPN on the frequency range [1 Hz – 50 kHz]. The total integrated RMS CEP noise is 325 mrad for an acquisition time of 1 s. This measurement confirms the single shot in-loop measurement over the full noise bandwidth. To our knowledge this is the first time that a spectral f-to-2f measurement is performed at 100 kHz.

Last, we report on intensity fluctuations of the complete laser source, which are known to be correlated with CEP noise through various amplitude to phase noise transfer mechanisms [29, 30]. As an example, self-phase modulation in the white light generation (inside the OPA and the f-to-2f), amplification stages, and MPC could cause such transfer. We characterize our system in terms of relative intensity noise. The measurement setup consists in a Si photodiode, a low-pass filter at Nyquist frequency and an oscilloscope. The RIN PSD and integrated RMS RIN (IRIN) measured after the MPC are shown in Figure 5(b). The total IRIN is 0.32% in the [1Hz – 50 kHz] bandwidth. We also point out a RIN spectral peak at 100 Hz that is clearly correlated to a peak at the same frequency in the CEP PSD (see Figure 5(a)). This is a clear evidence of an amplitude to phase noise transfer. As expected, a low value of IRIN is therefore paramount in the design of such CEP-stable system.

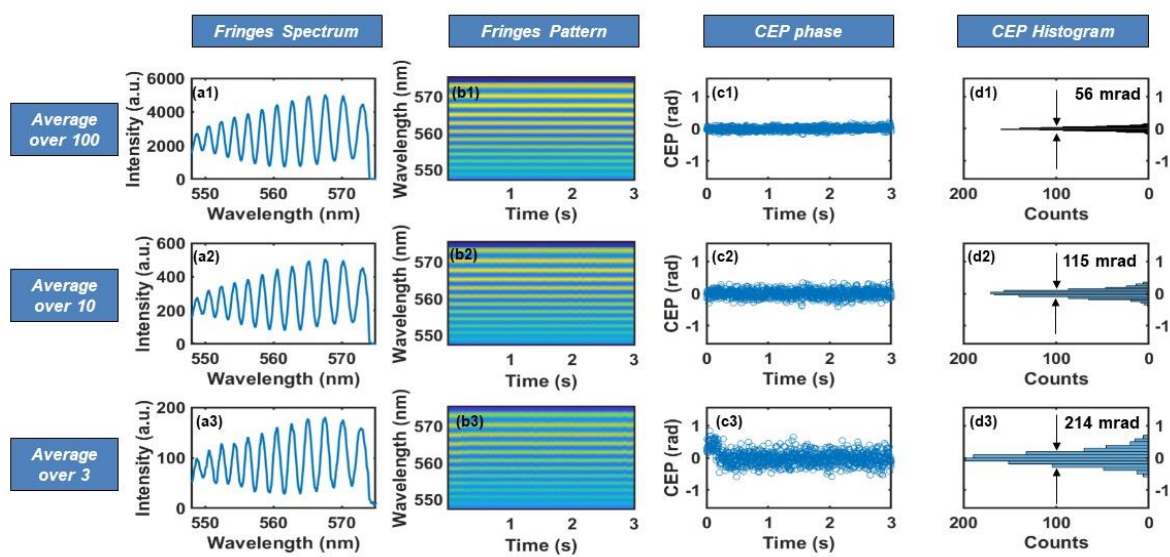

*Fig. 4. Out-of-loop f-2f spectral fringes measurement with a spectrometer. (a1-a3) Single acquisition of spectral fringes with integration times corresponding to 100, 10 and 3 pulses respectively. (b1-b3) Corresponding fringe pattern evolution as a function of time over 3 s. (c1-c3) Corresponding CEP drift as a function of time. (d1-d3) Histograms of CEP fluctuations.*

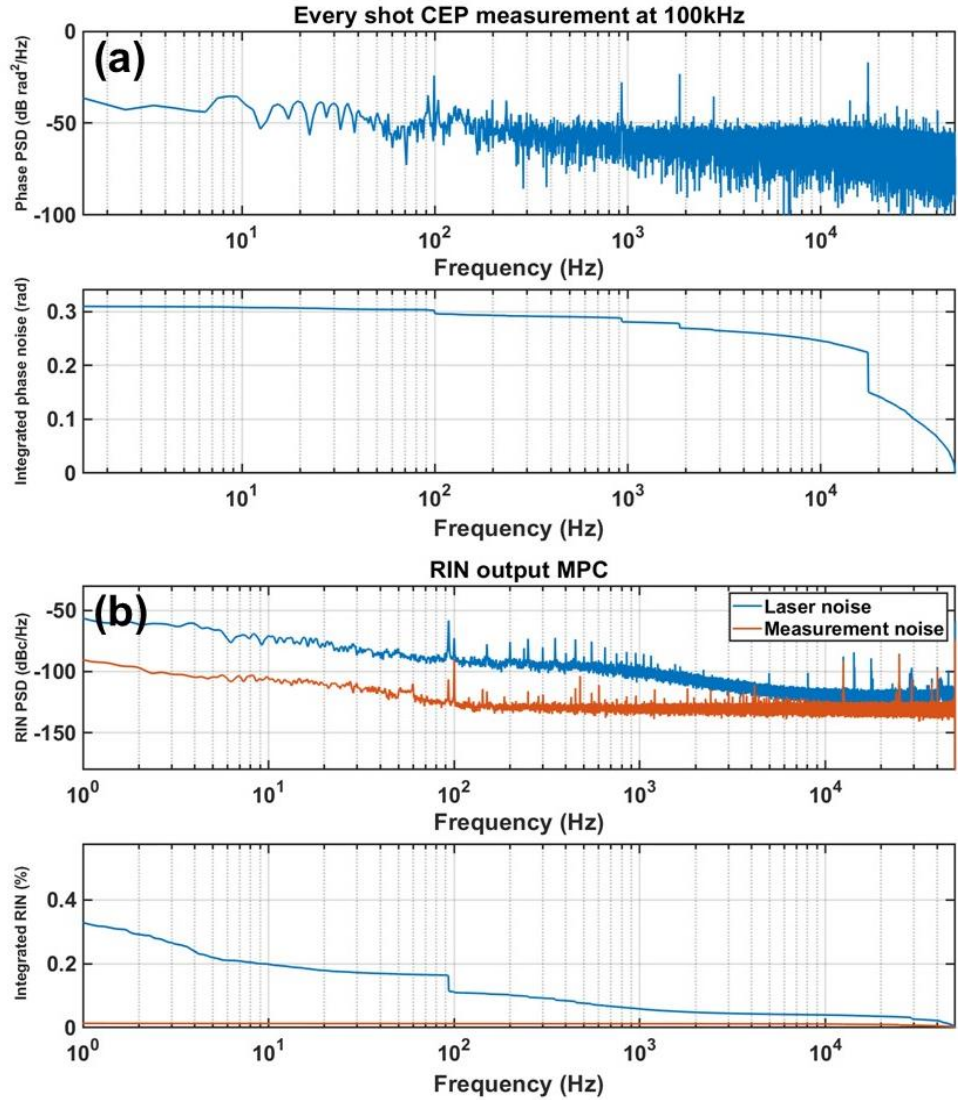

*Fig. 5. (a) Full PSD and IPN from single-shot, every-shot measurement. (b) PSD and IRIN at the output of the laser source.*

In conclusion, we demonstrated the CEP-stabilization of an ultrafast FCPA followed by an MPC-based nonlinear temporal compression stage. This proof-of-principle system delivers 30 µJ 96 fs pulses centered at 1030 nm, at 100 kHz repetition rate with a full-bandwidth single-shot CEP stability better than 400 mrad and RMS RIN of 0.32%. It brings clear evidence of the compatibility of all subsystems, including stretcher/compressor units, single-mode fiber amplifiers, large mode area power amplifier, and nonlinear compression in an MPC, with CEP stabilization. Taking into account the already demonstrated high energy, high power [11] amplifier architectures as well as the high efficiency nonlinear compression schemes [31], the system performances could realistically be scaled. We believe that this study paves the way to the development of compact, robust, CEP-stable, high-power (>100 W), high-energy (>500 µJ), few cycles sources particularly attractive for a wide range of applications such as XUV generation through HHG, attoscience, coincidence spectroscopy and nanoscale imaging.

**Funding.** European Union's Horizon 2020 research and innovation programme H2020-MSCA-ITN-2014-641789-MEDEA ; Agence Nationale de la Recherche (ANR-10-LABX-0039-PALM, ANR-16-CE30-0027-0) ; Conseil Départemental de l'Essonne (ASTRE Sophie).